\documentclass[aps,twocolumn,showpacs]{revtex4}

\usepackage[T1]{fontenc}
\usepackage{graphicx}
\usepackage{color}
\newcommand{\ket}[1]{|#1\rangle}
\newcommand{\bra}[1]{\langle #1|}

\pacs{03.75.Be,03.75.-b,39.25.+k}

\begin{document}
\title{Limitations of the modulation method to smooth a
wire guide roughness}

\author{I. Bouchoule, J.-B. Trebbia, and C. L. Garrido Alzar}
\affiliation{
Laboratoire Charles Fabry de l'Institut d'Optique,
CNRS et Université Paris 11, 91127 Palaiseau Cedex, France} \today
\begin{abstract}
It was recently demonstrated that wire guide roughness can be
 suppressed by modulating the wire 
currents~[Phys. Rev. Lett. {\bf 98}, 263201 (2007)]
so that the atoms experience a time-averaged 
potential without roughness.
In this paper, we  theoretically study the limitations of this technique.
At low modulation frequency,
we show that the longitudinal
potential modulation produces  heating of the cloud and we 
compute the heating rate. We also give a quantum derivation of the 
rough conservative potential associated with the micro-motion of the atoms.
At large modulation frequency, we compute the loss rate due to 
non adiabatic spin-flip and show that it presents resonances at 
multiple modulation frequencies.
These studies show that the modulation technique works for a wide range 
of experimental parameters.
We also give conditions to realize radio-frequency evaporative cooling
in such a modulated trap.
\end{abstract}

\maketitle

\section{Introduction}
 Atom-chips are a very promising tool for cooling and
manipulating cold atoms~\cite{Folmanrevue}.
 Diverse potentials, varying on the micron-scale, can be
realized and very high transverse
confinements are possible.
Envisioned applications range from integrated guided atomic
interferometry~\cite{Cornellchipinter,Schmiedmayerdoublepuits,Zimmermanninter}
to the study of low dimensional gases~\cite{esteve2006,trebbia2006,JacobTonks}.
 To take benefit from the atom chip technology, the atoms should be
brought close to the current carrying wires. But the atoms then
experience a rough potential due to wire
imperfections~\cite{Lukin-frag2003,Nous-fragm} and
this used to constitute an important limitation of the atom-chip
technology.
However,  a method to overcome this roughness problem,
 based on modulated currents,
was recently demonstrated~\cite{modulationexp}.
 Due to the important envisioned applications of this method, a
study of its limitations is crucial.

The method to suppress  atomic  wire guide  roughness 
consists in a fast modulation of the wire current 
around zero
so that the atoms, as in a Time Orbiting Potential (TOP) trap~\cite{top}, 
experience the time-average potential.
 Since the longitudinal potential roughness
is proportional to the wire current~\cite{Zimmermann-frag2002}, 
the time averaged potential is exempt from roughness.
The modulation frequency $\omega$
must be large enough so that the
atomic motion cannot follow the instantaneous potential.
 On the other hand,
$\omega$ should be small enough in order to prevent losses due
to spin-flip transitions~\cite{noteTOP}.
 In this paper, we present analysis that go beyond the 
time-averaged potential approach and we identify 
the limitations of this method, both 
for low and large $\omega$.
 We also investigate
the possibility of using the radio-frequency evaporative 
cooling method in such a modulated trap.

 In Sec.~II, we present the considered situation.
In Sec.~III, we investigate the limitations of 
the method that arise at small modulation frequency.
Using a Floquet analysis, we show that the atomic cloud is submitted to 
a heating that we quantitatively study.
Within this formalism, we also recover the well-known adiabatic 
potential experienced by atoms in fast modulated fields.
In our case, it amounts for a residual roughness.
 In sec.~IV, we compute the expected spin-flip
losses due to the time modulation of the magnetic
field orientation, that arise at large $\omega$.
 Finally, the last section gives some insights on the
possibility to realize radio-frequency evaporative cooling in the
modulated guide.

\section{Wire guide}
\begin{figure}
\includegraphics{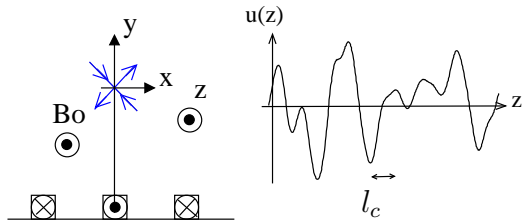}
\caption{A wire guide produced by three current carrying wires.
Geometrical deformation of the wire produce a longitudinal
potential roughness (of correlation length $l_c$)
 proportional to the wire current
as depicted in the figure. 
}
\label{fig.shemaguide}
\end{figure}

 A wire guide can be obtained combining a transverse quadrupolar field 
and a homogeneous longitudinal magnetic field $B_0$. 
The quadrupolar field can be realized using for example 
three current  carrying wires as
shown in Fig.\ref{fig.shemaguide}.
 Because of wire  deformations~\cite{Nous-fragm,Schumm2005frag},
the current density
inside the wires acquires non zero transverse components. 
This produces a longitudinal rough magnetic field $b_z$
proportional to the wire current, much smaller than the external 
field $B_0$.
The method to effectively remove the roughness consists
in modulating the  currents at
a frequency $\omega$ while the longitudinal field $B_0$ is kept
constant.

\section{Effect of the modulation on the longitudinal motion}
\label{sec.long}
 Let us first assume the modulation frequency  of the wire currents is
small enough so that the atomic spin orientation can
follow adiabatically the magnetic field orientation.
The atoms are then subjected to the instantaneous
potential $\mu |\bf{B}|$, where $\mu$ is the atomic magnetic 
dipole moment.
 For $B_0$ much larger than $E/\mu$ where $E$ is the typical
transverse atomic energy, the instantaneous transverse potential
is harmonic and proportional to the instantaneous wire currents.
Since the oscillation frequency of this potential
is modulated in time,
the transverse classical  dynamics
is described by  a
Mathieu equation, which predicts stable motion as long as
$\omega>0.87\omega_\perp$, where $\omega_\perp$ is the maximum
instantaneous transverse
oscillation frequency. This classical criteria is also
predicted by quantum mechanics since the Wigner function
evolves as a classical phase space distribution 
for a harmonic potential~\cite{PhysRevA.31.564}.
In this paper, we assume this stability condition is
fulfilled. We also assume the
longitudinal dynamics is decoupled from the transverse one 
and
we focus on the longitudinal motion.
 The longitudinal instantaneous potential is
\begin{equation}
V(z,t)= u(z)\cos(\omega t),
\end{equation}
where $u(z)=\mu b_z(z)$, sketched in Fig.\ref{fig.shemaguide},
is produced by wire deformations.
 The idea of the method to smooth the roughness
is that
the longitudinal motion of the atoms does not have time
to follow the time evolution of the potential.
The  atomic motion is then well described
by the effect of the conservative potential
$\langle V(z,t)\rangle$, where the time average is
done over a modulation period.
Since $\langle V(z,t)\rangle=0$, the atoms
do not experience any roughness.
 We study below the conditions on the modulation frequency $\omega$
for such an approach to be valid.

 As the Hamiltonian experienced by the atoms is periodic
in time, we use the well-known  Floquet
representation~\cite{Floquet},
briefly presented below for the situation 
considered here.
 A new quantum number
$n_F$ is introduced, 
which gives a relative
number of modulation energy quanta.
The Hamiltonian in this representation  is
time-independent  and contains two contributions.
The first one,
\begin{equation}
H_0=\sum_{n_F=-\infty}^{\infty}(p^2/(2m)+\hbar\omega
n_F)\ket{n_F}\bra{n_F},
\label{eq.H0chauff}
\end{equation}
does not couple different Floquet subspaces. The second one,
\begin{equation}
H_1=\sum_{n_F=-\infty}^{\infty}u(z)/2(\ket{n_F}\bra{n_F+1} + 
\ket{n_F+1}\bra{n_F}),
\label{eq.H2chauff}
\end{equation}
couples adjacent Floquet subspaces.
 If the state of the system in the Floquet representation is
$\sum_{n_F} \ket{\psi_{n_F}}(t)\ket{n_F}$, where
$\ket{\psi_{n_F}}(t)$ gives the state of the system
in the manifold of Floquet number $n_F$,
then the state of the system in the bare representation is
$\sum_{n_F} \ket{\psi_{n_F}}(t)e^{-in_F \omega t}$.
Expectation values of observables contain cross terms
involving different Floquet numbers. However,
 as long as evolution on time scales much larger than
$1/\omega$ is considered, such cross terms (interference terms)
 average to zero and the
different Floquet states can be interpreted as
physically different states.
 A given state has an infinite number of
Floquet expansions. In particular, it is possible
to assume that the
initial state is in the Floquet manifold of Floquet number
$n_F=0$.

Let us consider a state $\ket{p_0,n_F=0}$ of momentum $p_0$
in
the Floquet manifold $n_F=0$. The modulated rough potential $u$
is responsible for two different phenomena.
 First, it induces a change rate of the atomic energy. This
irreversible evolution is due to the continuous nature of the
rough potential Fourier decomposition:~the state
$\ket{p_0,n_F=0}$  is coupled
to a continuum of momentum states of the adjacent
Floquet subspaces $n_F=\pm 1$ and
 this coupling to a continuum induces
a departure rate from the initial state,
associated to a rate of kinetic energy change.
 Second, the modulated potential is responsible for
the well-known adiabatic potential experienced
by atoms in fast modulated fields~\cite{LandauVad}.
We show that this adiabatic potential is due to
processes of order two in $u$ that
couple the state $\ket{p_0,n_F=0}$ to
the states $\ket{p_1,n_F=0}$
{\it via}
the virtually populated intermediate states $\ket{q,n_F=\pm 1}$.

 In the first sub-section, we  investigate the
first phenomenon and  compute the associated heating
rate for a cloud at thermal quasi-equilibrium.
 In the second subsection, we derive the adiabatic
potential experienced by the atoms.
In both sections, we emphasize on the case where the
potential roughness is that obtained at large distance
from a flat wire having white noise border fluctuations.

\subsection{Heating of the atomic cloud}
\begin{figure}
\includegraphics{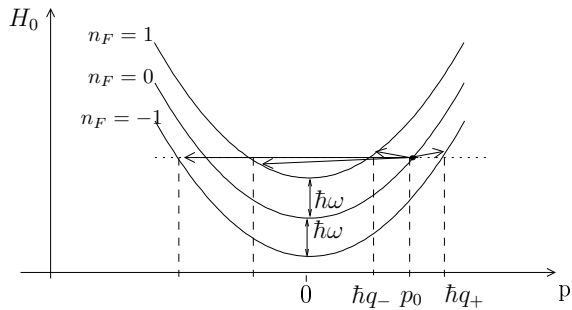}
\caption{Transitions responsible for a heating of the atomic cloud.
The parabolas give the energy $H_0$, given in Eq.~(\ref{eq.H0chauff}),
versus the momentum $p$ for different Floquet manifolds $n_F$. 
The state of momentum $p_0$ in the Floquet manifold
$n_F=0$ is
coupled to different momentum states of
the Floquet manifolds $n_F=\pm 1$  by the 
rough potential $u(z)$.
}
\label{fig.Floquet1}
\end{figure}

Let us suppose the atom is initially in the
state $\ket{p_0,n_F}$ of momentum
$p_0$ in the Floquet manifold
$n_F=0$.
As shown in Fig.\ref{fig.Floquet1},
this state is coupled by $u$ to the continuum of
momentum states in the Floquet manifold $n_F =\pm 1$,
which leads to
a decay of the initial state population.
 The momenta of the final states that fulfill energy
conservation in the Floquet subspace $n_F=-1$ are
$\pm\hbar q_{+}$ where $q_+=\sqrt{k_0^2+2m\omega/\hbar}$,
$k_0=p_0/\hbar$ being  the initial atomic wavevector.
Decay towards these states involves the Fourier component
$\pm q_+-k_0$ of $u$ and
increases the kinetic energy of the atom by $\hbar\omega$.
  If  $k_0^2>2m\omega/\hbar$,
there exist states
in the  Floquet subspace $n_F=1$
that have the same energy as the initial state.
The momentum of those final states  are
$\pm \hbar q_{-}$ where $q_{-}=\sqrt{k_0^2-2m\omega/\hbar}$
and decay towards these states
decreases the kinetic energy of the atom by $\hbar\omega$.
 A perturbative calculation, identical to the one used to
derive  Fermi Golden rule, gives an energy exchange  rate
\begin{equation}
\begin{array}[t]{l}
\frac{dE}{dt}=\frac{\pi\omega m}{2\hbar^2}
[ (S(-k_0 + q_{+})+S(-k_0-q_{+}))
/q_+\\
 -\Theta(|k_0|-\sqrt{2m\omega/\hbar})
(S(-k_0+q_{-})+S(-k_0-q_{-}))/q_- ]\\
\end{array}
\label{eq.dEdtquant}
\end{equation}
where $S(q)=1/(2\pi)\int e^{iqz}\langle u(0)u(z)\rangle dz $
is the spectral density of $u$, characterized by the
correlation length $l_c$, and 
$\Theta(x)$ is the Heaviside
function that is zero for $x<0$ and 1 for $x>0$. The derivation
of Eq. (\ref{eq.dEdtquant}) is detailed in the appendix A.
 As pointed out in the appendix, Eq. (\ref{eq.dEdtquant}) is
not valid for an initial momentum very close
to $\sqrt{2m\hbar\omega}$.
However, the range of $k_0$ for which the formula is not
valid is in general very small and we ignore this
in the following.

Apart from the rms amplitude of the roughness, which accounts
only for a multiplicative factor in the rate of the energy change,
three energies are relevant:
$E_\omega=\hbar\omega$ is the energy quantum corresponding to the
modulation frequency, $E_{m}=ml_c^2\omega^2$ is about the
kinetic energy of an atom that would
travel over $l_c$ during an oscillation period,
and $E_c=p_0^2/(2m)$ is the atomic kinetic energy.
 In the following, we consider two different limits
for which we give simplified expressions for the energy
change rate: the classical limit and the quantum low energy limit.

  Let us first assume that $E_\omega/\sqrt{E_m E_c}\ll 1$ and
$E_\omega E_m^{1/2}/E_c^{3/2}\ll 1$. We show below that this
two conditions ensures the validity of the classical 
behavior.
 This two conditions ensure that 
$E_\omega/E_c\ll 1$ so that
$q_+$ and $q_-$ are close to $k_0$
and one can expand the quantity $q_\pm /k_0$  in
powers of $m\omega/(\hbar k_0^2)$.
 Since the wavevectors $-k_0 -q\pm$ and 
$-k_0 +q\pm$ are separated by about $2k_0$, 
the first condition ensures that 
the spectral components $S(-k_0 -q\pm)$ are negligible
compared to the two other ones.
The second condition ensures that the latter are
well approximated using a Taylor expansion of $S$.
 Finally, the energy exchange rate writes
\begin{equation}
\frac{dE}{dt}= -\left [
2\omega^2/v_0^3 S(\omega/v_0)+S'(\omega/v_0)\omega^3/v_0^4
\right ]
\pi/(2m),
\label{eq.DEDTv0}
\end{equation}
  where $v_0=\hbar k_0/m$ is the atomic velocity.
 This energy exchange rate does not depends on $\hbar$ and is thus a classical
result. It is obtained through a classical calculation
of kinetic energy exchange computed after
expanding the atomic trajectory to second order in
$u$.
Note that, using the classical expression $E_c=mv_0^2/2$,
the two conditions
$E_\omega/\sqrt{E_c E_m}\ll 1$  and
$E_\omega E_m^{1/2}/E_c^{3/2}\ll 1$
 are verified in the limit where
$\hbar$ goes to zero, as expected for the validity of classical
physics.

 Let us now consider the limit
$E_\omega/E_c\gg 1$
and $E_c E_m^{1/2}/E_\omega^{3/2}\ll 1$, that
we denote the quantum low energy limit.
The first inequality ensures that the Heaviside function
in Eq. (\ref{eq.dEdtquant}) is zero and that
$q_+$ can be replaced by $\sqrt{2\hbar\omega}$
in the denominator.
The second inequality ensures that this
replacement
is also valid for the argument of the $S$ function.
Then the
energy exchange rate given by Eq. (\ref{eq.dEdtquant}) reduces to
\begin{equation}
\frac{dE}{dt}=\frac{\pi\sqrt{m\omega}}{\sqrt{2\hbar^3}}
\left [S(-k_0+\sqrt{2m\omega/\hbar})
+S(k_0+\sqrt{2m\omega/\hbar})\right ].
\label{eq.dEdtquantlowE}
\end{equation}
 This is a
quantum result, sensitive to the fact that energy exchange
between the atom and the oscillating potential involves
the energy quanta $\hbar \omega$. In the limit where
$E_c\ll E_\omega^2/E_m$ ($k_0\ll 1/l_c$), it converges towards a finite value
\begin{equation}
\frac{dE}{dt}=\frac{\pi\sqrt{2m\omega}}{\sqrt{\hbar^3}}
S(\sqrt{2m\omega/\hbar}),
\label{eq.dEdtlim}
\end{equation}
that does not depends on the initial momentum $\hbar k_0/m$.

 Let us now consider a cloud initially at thermal equilibrium with a velocity
 distribution $n(v_0)$.  The heating rate,
obtained after averaging Eq. (\ref{eq.dEdtquant}) over $n(v_0)$, is
\begin{equation}
k_B\frac{ dT}{dt}=2\int_{0}^{\infty} n(v_0) \frac{dE}{dt} dv_0 \label{eq.dTdt},
\end{equation}
where $k_B$ is the Boltzmann factor.
Although the heating rate depends on the precise shape of
the spectral density $S$, some general properties can  be derived.

 First, although the energy exchange rate may be negative
for some velocities, we show below that $dT/dt$ is always positive.
For a longitudinally homogeneous gas, this positivity
ensures the increase of the entropy, as required by the
second law of thermodynamics in the absence of heat exchange with the
cloud and without gaining information on the system.
To demonstrate that  $dT/dt>0$,
we perform a change of variables in the four integrals
obtained by substituting Eq.~(\ref{eq.dEdtquant}) into 
Eq.~(\ref{eq.dTdt})
to find
\begin{equation}
\renewcommand{\arraystretch}{1.5}
\begin{array}[t]{l}
 k_B\frac{d T}{dt}=\pi\omega\sqrt{m}/\hbar\\
\int_0^{Q_0} \frac{dq}{q}S(q)
\left (
n(\omega/q-\hbar q/(2m))-n(\omega/q+\hbar q/(2m))\right )\\
+
\int_{Q_0}^{\infty} \frac{dq}{q}S(q)
\left ( n(\hbar  q/(2m)-\omega/q)-n(\omega/q+\hbar q/(2m))\right )\\
\end{array}
\label{eq.pos}
\end{equation}
where $Q_0={\sqrt{2m\omega/\hbar}}$.
For a thermal equilibrium distribution, $n(v)$ is a decreasing
function of $|v|$. Furthermore, the spectral density is a positive
function.
We thus find that  $dT/dt>0$ so that the effect of the potential roughness
is always a heating of the cloud.
Eq.~(\ref{eq.pos}) also shows that the heating rate goes to zero at very
large temperatures, since $n$ is then about flat over the explored
velocities.

Second, for large enough temperatures one expects to recover the classical result
and the heating rate should not depend on $\hbar$. Then the heating rate depends
only on the four independent quantities  $\langle u^2\rangle$, $E_m$, $\omega$ and
$k_B T$. Since  $\langle u^2\rangle$ enters only as a multiplicative factor in the
heating rate, using dimensional analysis 
we show that $k_B dT/dt$ is the product of
${\langle u^2\rangle}/{(m\omega l_c^2)}
$ and
a function of $k_B T/E_m$. As a consequence, if the function
giving the heating rate versus $T$ is known for a given
value of $\omega$ and $E_m$,  then the heating rate is known
for any value of $T$, $E_m$ and $\omega$.

Finally, at low enough temperatures,
the  heating rate is well estimated by
substituting  Eq.~(\ref{eq.dEdtquantlowE})
into Eq.~(\ref{eq.dTdt}).
 One  expects that the heating rate converges
towards Eq.~(\ref{eq.dEdtlim}) when the 
temperature becomes much smaller than
$\hbar\omega$ and $\hbar^2/(ml_c^2)$.

\begin{figure}
\hspace*{-1cm}{\includegraphics{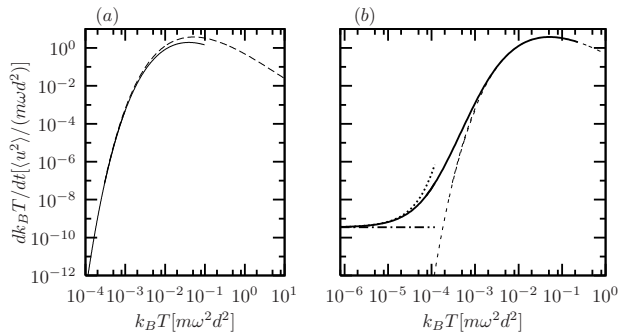}}
\caption{Heating rate of a cloud as a function of
its temperature, for a modulated rough
potential whose spectral density is given by Eq.~(\ref{eq.Ju}).
Figure (a): classical predictions (dashed lines) and asymptotic
behavior  at low temperature given by Eq.~(\ref{eq.dTdtxgrand}) (solid line).
Figure (b): exact result for $\omega/(2\pi)=200\hbar/(md^2)$ (solid
line) compared with the  quantum low energy prediction (dotted line),
the asymptotic prediction of Eq.~(\ref{eq.dTdtlimitJu})
(dashed-dotted lines) and the classical result (dashes).
}
\label{fig.dTdt}
\end{figure}

In the following, we give quantitative results in the case of
a potential  roughness  obtained at
large distances $d$ from a flat wire whose borders have white noise
fluctuations of spectral density $J_f$.
In this condition, the  spectral density
of $u$ is given by~\cite{Lukin-frag2003,Nous-fragm}
\begin{equation}
S(k)=J_f\frac{(\mu_0\mu I)^2}{4\pi^2}k^4K_1(kd)^2,
\end{equation}
where $K_1$ is the modified Bessel function of the first kind.
The typical correlation length of $u$ is the distance above the wire
$d$ so that $E_m=m\omega^2d^2$. The mean square of the rough potential is
$\langle u^2\rangle\simeq 0.044(\mu\mu_0 I)^2J_f/d^5$.
In the following we use $\langle u^2\rangle$ as a parameter
instead of $J_f$.
 The spectral density
of $u$ is then 
 \begin{equation}
S(k)=\alpha \langle u^2\rangle d (kd)^4K_1(kd)^2 ,
\label{eq.Ju}
\end{equation}
where $\alpha \simeq 23$.

 We first study the heating rate predicted by  classical
physics. This classical heating rate
is plotted in Fig.\ref{fig.dTdt}(a) as a dashed line.
The temperature and heating rate are scaled to $E_m$ and
$\omega\langle u^2\rangle/E_m $
respectively so that the curves corresponding to the
classical predictions are independent of the problem parameters.
 We observe the expected decrease to  zero of the heating rate at
high temperatures.
 We also observe a rapid decrease of the heating rate
as the temperature decreases, for temperatures  much smaller than $E_m$.
 The maximum heating rate is about
$
2.1 \langle u^2\rangle /(m\omega d^2)$ and
is obtained for the temperature $k_B T_M\simeq 0.07 E_m$.
 For $d=5~\mu$m and $\omega/(2\pi)=50~$kHz, which are
parameters similar to that of the experiment presented in~\cite{modulationexp},
$T_M=1.8~$mK. Typical
cold atoms temperatures are much smaller than this value and it
is thus of experimental interest to investigate in more detail
the regime $T\ll T_M$.

 The decrease of the heating rate  for $T\ll T_M$ is expected since,
in this case,
the atoms move on a distance much smaller than
the correlation length of the rough potential
during a  modulation period. The atoms are then
locally subjected to an oscillating force almost independent of $z$
and  the atomic motion can
be decomposed into a fast micro-motion in counterphase with the
modulation and a slow motion. The micro-motion being almost in counterphase 
with
the excitation force, almost no energy exchange between the atom and the
potential arises on a time scale larger than the modulation period.
 More quantitatively,  we can derive an analytical expression of the heating
rate in the regime where $T\ll T_M$, which shows
the decrease of the heating rate as temperature
decreases.
 For such low temperatures, as shown  {\it a posteriori} below,
wavevectors in $S$ that contribute to the heating rate
are much larger than $1/d$ so that  we can replace
the Bessel function $K_1(x)$
in Eq.~(\ref{eq.Ju})
by its asymptotic value at large $x$.
We then find that the integrand in Eq.~(\ref{eq.dTdt})
is peaked around $v_0=2^{1/3}(k_B T\sqrt{E_m})^{1/3}/m$ and
the Laplace method
gives the following approximation  for the heating rate:
\begin{equation}
\frac{k_B dT}{dt}
=\beta \frac{\omega \langle u^2\rangle}{E_m}
\left (\frac{E_m}{k_B T}\right )^{7/3}e^{-3(E_m/(2k_B T))^{1/3}},
\label{eq.dTdtxgrand}
\end{equation}
where $\beta \simeq 0.36$.
 This asymptotic function is plotted in Fig.\ref{fig.dTdt} (a)
(solid line). It coincides with the exact classical result
within 20\% as long as $k_B T< 0.002 E_M$.
The above expression of $v_0$ and Eq.~(\ref{eq.DEDTv0}) validate
the expansion at large $x$ of the Bessel function $K_1(x)$
for $T\ll T_M$.

 The limit of validity of the classical results described above
is given by $E_\omega/\sqrt{E_c E_m}\ll 1$ and
$E_\omega E_m^{1/2}/E_c^{3/2}\ll 1$, where
$E_c\simeq m v_0^2$, $v_0$ being the typical velocity
involved in the heating process.
Using the above value for $v_0$, the condition
of validity of the classical regime reduces, for
$E_\omega\ll E_m$, to  $k_B T \gg E_\omega$.
For  $\omega/(2\pi)=50~$kHz, we  find that the classical regime
fails for temperatures $T \ll 2~\mu$K. At lower temperatures,
quantum analysis is required to estimate the heating rate.

For the above parameters 
($d=5~\mu$m and $\omega/(2\pi)=50~$kHz),
$E_m/E_\omega= 10^{4}$ and  the heating rate is exponentially
small at temperatures smaller than  a micro-Kelvin where
classical physics fails
(the  term $e^{-3(E_m/(2k_B T))^{1/3}}$ 
in Eq.~(\ref{eq.dTdtxgrand}) is $3\times 10^{-31}$ for
$T=1~\mu$K.).
 Thus,
in order to investigate the heating rate beyond the classical approximation,
 we consider a different situation for which
$E_m/E_\omega$ is only equal to 200. This would correspond, for
the same distance $d=5~\mu$m, to a modulation frequency of only
1~kHz.
 The exact heating rate, which is computed by
substituting Eq.~(\ref{eq.dEdtquant}) into
Eq.~(\ref{eq.dTdt}), is plotted in Fig.\ref{fig.dTdt} $(b)$. This
calculation shows that the classical result is valid up to a
factor of 2 as long as $k_B T>0.2E_\omega$. At  lower
temperatures, the classical result underestimates
the heating rate. At temperature much smaller than
$E_\omega^{3/2}/\sqrt{E_m}$,
the heating rate is well approximated by
the predictions
in the low energy quantum limit where Eq.~(\ref{eq.dEdtquantlowE}) is valid.
This prediction is represented as a dotted line in the graph.
At temperatures much smaller than $E_\omega^2/E_m$
({\it i.e.} for $k_B T\ll \hbar^2/(md^2)$), the heating rate
converges towards Eq.~(\ref{eq.dEdtlim}).
Assuming $E_m/E_\omega\gg 1$, then the expansion of $S$
at large wavevector can be used and Eq.~(\ref{eq.dEdtlim}) gives
\begin{equation}
k_B dT/dt
=\zeta \langle u^2\rangle /\hbar  ( m\omega d^2/\hbar )^2e^{-2\sqrt{m\omega d^2/\hbar}}
\label{eq.dTdtlimitJu}
\end{equation}
where $\zeta \simeq 4.0$.
 This asymptotic value is plotted in Fig.\ref{fig.dTdt} as
 dashed-dotted lines.
The heating rate is equal to this limit up to a factor of 2 as soon as
$k_B T< 0.2 E_\omega^2/E_m$.

 The heating of the atomic cloud
can easily be made small enough experimentally to have no noticeable
effects.
 Let us for example consider the situation, similar to the
experiment in~\cite{modulationexp}, where
 $d=5~\mu$m and $\sqrt{\langle u^2\rangle}=50~$nK.
If the modulation frequency is as low as 1~kHz, then
the maximum heating rate is 3~$\mu$K/s and is obtained for a temperature of
700~nK. Thus, for such a low modulation frequency, 
the heating may be a problem in experiments using the 
modulation technique.
 However, as soon as the modulation frequency is increased to
50~kHz, as in~\cite{modulationexp},
the maximum heating rate is only
of 64~nK/s and is obtained at a large temperature of $1.8~$mK. At lower
temperature, the exponential decrease of the heating rate shown in
Eq.~(\ref{eq.dTdtxgrand}) rapidly decreases the heating 
rate to completely negligible values.

\subsection{Effective remaining potential}
 In this subsection, we show that Raman processes (of second order in $u$),
in which adjacent Floquet states are virtually populated,
are responsible for an effective potential
\begin{equation}
V_{ad}=\left (\partial u/\partial z \right )^2/(4m\omega^2).
\label{eq.Had}
\end{equation}
 This potential is a well-known classical result~\cite{LandauVad} that
corresponds to the kinetic energy of the micro-motion
of a trapped particle.
The micro-motion has been seen, for example, in Paul 
traps~\cite{paul-micro,paul-ions} 
and in TOP traps~\cite{top-micro-1,top-micro-2}.
In our situation, at large oscillation frequency, the micro-motion has an amplitude
$\xi \simeq -(\partial u/\partial z) 
/(m\omega^2)\cos(\omega t)$ much smaller than the
correlation length of $u$. It is in counterphase with the excitation force
and has a kinetic energy $V_{ad}$. 
In this limit, since the micro-motion 
is in counterphase with the excitation force, the 
energy transfer between the atom and
the potential, averaged over a modulation period,  vanishes. 
Energy conservation then shows that  
the slow motion of the atom is subjected to
an effective rough potential $V_{ad}$.
 It is well-known that
this effective potential, due to the fast atomic micro-motion,
is responsible for the confinement in rapidly modulated
Paul traps.
 A quantum derivation of $V_{ad}$
has already been done in~\cite{PhysRevA.31.564} using
a secular approximation. Here we give an alternative
derivation based on the Floquet representation.

\begin{figure}
\includegraphics{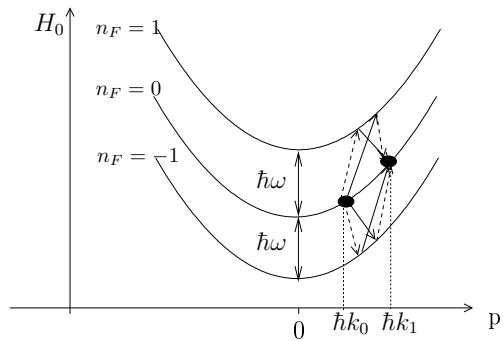}
\caption{Raman transitions responsible for the
adiabatic potential of Eq.~(\ref{eq.Had}).
The second order coupling between two momentum states
of wavevector $k_0$ and $k_1$ is sketched.
The coupling
produced by two Fourier components
of $u$ of wavevector $q$ and $q'=k_1-q-k_0$  are represented
as dashed and solid arrows respectively.
}
\label{fig.Had}
\end{figure}

 Let us   compute the effective coupling between the states
$\ket{k_0}$ and $\ket{k_1}$
of momentum $\hbar k_0$ and $\hbar k_1$ respectively, both being in
the Floquet subspace $n_F=0$.
 For this purpose, we first
investigate the effect of a given pair of Fourier components
of $u$ that couple the two previous states. Their wavevectors
are $q$ and $q'=k_1-k_0-q$.
Four processes are involved in the effective coupling between
$\ket{k_0}$ and $\ket{k_1}$, as sketched in Fig.\ref{fig.Had}, and the
effective coupling is the sum of the four amplitudes.
 The precise effective coupling between two ``ground'' states
coupled via an intermediate level has been investigated
in~\cite{KlausRaman}. The authors show that
the effective coupling is $V_1V_2/\Delta$, where
$V_1$ and $V_2$ are the coupling to the intermediate
state and $\Delta$ is the difference between the
energy of the intermediate  state and the mean energy
of the two  ``ground'' states.
  Using this result,
we find that the effective coupling associated with each
process is
\begin{equation}
v_{\rm{eff}}=u_q u_{q'}/(4(\hbar^2(k_0+\kappa)^2)/(2m)\pm\hbar\omega-E_0),
\end{equation}
where $\kappa$ is $q$ or $q'$ depending on the
process and $E_0=\hbar^2 (k_0^2+k_1^2)/(4m)$.
Adding the four amplitudes, we find
\begin{equation}
V_{\rm{eff}}=\frac{u_q u_{q'}}{4}
\begin{array}[t]{l}
\left ( \displaystyle\frac{\hbar^2(k_0+q)^2/m -2E_0}{(\hbar^2(k_0+q)^2/(2m) -E_0)^2
-\hbar^2 \omega^2} \right .\\
+\left . \displaystyle\frac{\hbar^2(k_0+q')^2/m -2E_0}{(\hbar^2(k_0+q')^2/(2m) -E_0)^2
-\hbar^2 \omega^2 } \right ).\\
\end{array}
\end{equation}
 Assuming that the kinetic energies of the final, initial and 
intermediate states are all much
smaller than $\hbar \omega$, the denominator can be simplified
to $\hbar^2\omega^2$ and we obtain
 \begin{equation}
V_{\rm{eff}}=\frac{u_q u_{q'}q q'}{2m\omega^2}.
\end{equation}
Doing the sum over the pairs ($q,q'$), we find
that the coupling between the momentum states
is that realized by the potential of Eq.~(\ref{eq.Had}).

 An alternative way to derive the adiabatic potential is 
to use a dressed Floquet representation. As shown in 
 appendix B, the calculations are more straightforward 
in this representation. In addition,  no detailed knowledge of the 
effective coupling corresponding to a transition
through a virtually populated state is required.

 The residual roughness given in
Eq.~(\ref{eq.Had}) constitutes a limitation of the
modulation method.
 It scales as $\left <u^2\right > /( ml_c^2\omega^2)$ where
$l_c$ is the typical correlation
length of $u$. Thus  it is much smaller than the
initial roughness amplitude as soon as
$\sqrt{\left <u^2\right >}\ll ml_c^2\omega^2$.
 In the case where
the roughness potential spectral density is that obtained at
large distances $d$ from a wire having white noise border
fluctuations of spectral density $J_f$,
we  obtain a mean value
\begin{equation}
\langle V_{\rm{eff}}\rangle =0.048 J_f
\frac{(\mu_0\mu I)^2}{m\omega^2d^7}=1.1\langle u^2\rangle /(m\omega^2d^2).
\end{equation}
 If the wire edges deformations have
a gaussian probability distribution, then
the roughness of the remaining potential is simply
$\sqrt{\langle V_{\rm{eff}}^2\rangle -\langle V_{\rm{eff}} \rangle
  ^2}=
\sqrt{2}\langle V_{\rm{eff}}\rangle$.
For $d=5~\mu$m, $\omega/(2\pi)=50~$kHz and
$\sqrt{\langle u^2\rangle}/k_B=500~$nK, the roughness of the effective
remaining potential is as small as 0.09~nK.

\section{Losses due to spin-flip transitions}

 All the previous analysis assume the atomic spin can  adiabatically
follow
 the direction of the instantaneous field when the current is
modulated. In this section, we investigate the conditions on the modulation
frequency for this adiabatic following requirement 
to be valid. Non adiabaticity
induces losses via spin-flip transitions to the untrapped states. 
Intuitively, we
expect that the losses are small for a modulation frequency 
much smaller than the
Larmor frequency. 
In the following we compute this loss rate,
following calculations done in~\cite{PhysRevA.56.2451} for the DC-case.

 Let us consider a  spin-one atom in the modulated guide
described in section II with $\omega\gg \omega_{\perp}$.
 We choose a coordinate system  whose origin is at the quadrupole field
center and of axis $x$ and $y$ at 45$^{\rm o}$ with the quadrupole
axis as depicted in Fig.\ref{fig.shemaguide}. 
We assume the atomic magnetic 
moment is $\mu {\bf J}/\hbar$ where ${\bf J}$ is 
the atomic spin angular momentum of components 
$J_x,J_y$ and $J_z$ along $x,y$ and $z$, respectively.
We note $\ket{\pm 1}$ 
and $\ket{0}$ the eigenstates of 
$J_z$ of eigenvalues $\pm\hbar$ and 0, respectively. 

 The magnetic field direction depends on the position
and on time.
 We apply a spatially and time dependent spin rotation
${\cal R}(t,x,y)$ so that ${\cal R}\ket{1}$ points along 
the instantaneous local
magnetic field direction. In such a  representation, the Hamiltonian is
\begin{equation}
H={\cal R}^{-1}\frac{ {\bf p}^2}{2m}{\cal R} + U
J_z/\hbar +i\hbar\frac{d{\cal R}^{-1}}{dt}{\cal R} \label{eq.Hamtrans}
\end{equation}
where 
\begin{equation}
  U=\mu B_0+m\omega_\perp^2\cos^2(\omega t) (x^2+y^2)/2.
\end{equation}
Here we assume $m\omega_\perp^2(x^2+y^2)\ll \mu B_0$ so that 
the harmonic approximation is valid. We also neglect the
effect of  gravity. This later assumption is relevant as soon as $g\ll
l_\perp\omega_\perp^2$ where $l_\perp=\sqrt{\hbar/(m\omega_\perp)}$ 
is the harmonic
oscillator length. 
This condition ensures that the time-averaged potential is barely
affected by the gravity and that 
the acceleration of the atoms over the spatial extension
of the trapped state has a negligible effect.

 We choose the rotation  $\cal R$ as a product of a rotation along $x$ 
and a rotation along $y$.
 To first order in $x$ and $y$, $\cal R$ is 
\begin{equation}
{\cal R} = 1 +i\theta_x J_x/\hbar + i \theta_y J_y/\hbar,
\end{equation} 
where the angles of the rotations along 
$x$ and $y$ are  $\theta_x=-{b'x\cos(\omega t)}/{B_0}$ and
$\theta_y={b'y\cos(\omega t)}/{B_0}$, respectively.
Here $b'=\omega_{\perp}\sqrt{2mB_0/\mu}$
is the quadrupole gradient at maximum current.
Calculation to first order in $\theta_x,\theta_y$ gives
\begin{equation}
{\cal R}^{-1}\frac{{\bf p}^2}{2m}{\cal R}= \frac{{\bf p}^2}{2m} +V_k
\end{equation}
where 
\begin{equation}
V_k=\frac{\sqrt{2}\hbar b'\cos(\omega t)}{m B_0}(p_x-ip_y)\ket{0}\bra{1}
+h.c.
\label{eq.Vk}
\end{equation}
and $h.c.$ stands for hermitian conjugate.
Here, we ignore the state $\ket{-1}$, which is relevant 
for low enough coupling (see below).
 The term $V_k$, due to the fact that ${\cal R}$ depends
on the position, is responsible for spin-flip losses in time independent Ioffe
magnetic traps~\cite{PhysRevA.56.2451}. 
  Within the approximations made here, 
the position dependence of $\cal R$ 
has no effect on the Hamiltonian within the spin state $\ket{1}$  
manifold: the Coriolis coupling analyzed in~\cite{Ho}, that 
corresponds to a rotation frequency proportional to $b'^2$, 
is not seen in this calculation.

Similar calculations give
\begin{equation}
i\hbar\frac{d{\cal R}^{-1}}{dt}{\cal R}= 
\frac{\hbar \omega  b'}{\sqrt{2}B_0}\sin(\omega t)
(x-iy)\ket{0}\bra{1}+h.c. \label{eq.dRdt}
\end{equation}
This term, due to the  time modulation of the local spin orientation, may also
produce spin-flip losses in modulated traps. The condition $\omega\gg
\omega_{\perp}$ ensures that the term of Eq.~(\ref{eq.dRdt}) has an effect much larger
than that of Eq.~(\ref{eq.Vk}) and we neglect the latter in the following.

 As in the previous section, we use the Floquet representation.
The Hamiltonian $H$ in the manifold of spin state $\ket{1}$ is
decomposed 
into the term
\begin{equation}
H_{0}=\sum_{n_F=-\infty}^{\infty}(p^2/2m + m\omega_\perp^2(x^2+y^2)/4+
n_F \hbar\omega)
\ket{1,n_F}\bra{1,n_F} \label{eq.Hadtrans}
\end{equation}
and the term
\begin{equation}
H_2=m\omega_\perp^2(x^2+y^2)/4\left ( \sum_{n_F=-\infty}^{\infty} \ket{1,n_F+2}\bra{1,n_F} + h.c. \right )
\label{eq.H2spinflip}
\end{equation}
that couples the Floquet $n_F$ state
 to the Floquet states $n_F\pm 2$. Here, $\ket{1,n_F}$ is the state vector of an atom
 in the spin state $\ket{1}$ with the Floquet number $n_F$.
The term $H_2$ is due to the part of Eq.~(\ref{eq.Hamtrans}) that is proportional to
$\cos(2\omega t)$.
 Since we assumed $\omega\gg\omega_\perp$, the effect of $H_2$
is weak and can be treated perturbatively.

We will compute the loss rate of an atom initially in the spin state $\ket{1}$ of
Floquet number $n_F =0$ and in the ground state ${\phi_0}$ of $H_0$.
 The term of Eq.~(\ref{eq.dRdt}) couples this trapped state to
the untrapped spin states $\ket{0}$ of Floquet numbers $n_F =\pm 1$.
  The energy spectrum of the spin state $\ket{0}$, which is unaffected 
by the magnetic
field, is a continuum. Coupling to this continuum leads to a
departure rate from the initial state, provided the Markov
approximation is fulfilled~\cite{Markov}. This approximation also 
ensures that the state $\ket{-1}$ can be neglected.
 We will show below that  
losses to the Floquet manifold
$n_F=+1$ are much larger than losses to the Floquet manifold $n_F=-1$.
Thus, we
consider in the following the final 
states in the manifold $n_F=+1$. 
Since $i\hbar
\frac{d{\cal R}^{-1}}{dt}{\cal R}$ does not
affect the longitudinal motion, 
we concentrate on the transverse degrees of freedom
and normalize $\phi_0$ as $\int\!\!\int\!\! dx dy |\phi_0|^2=1$. In
addition, because $i \hbar \frac{d{\cal R}^{-1}}{dt}{\cal R}\phi_0$ 
is, up to a 
phase factor $e^{i\theta}$, invariant under rotation 
of angle $\theta$ in the $xy$ 
plane, the losses to spin 0 states  are isotropic in the $xy$ plane. 
It is thus
sufficient to compute  the departure rate towards a plane wave 
travelling in the $x$
direction. The final state wavevector is 
\begin{equation}
k_f=\sqrt{2m(\mu B_0
+\hbar\omega_\perp/\sqrt{2}-\hbar\omega)}/\hbar,
\end{equation} 
and the Fermi Golden rule gives the
departure rate
\begin{equation}
\Gamma_0= \sqrt{2}\pi
\frac{\omega^2}{m \mu B_0 \omega_\perp} \hbar k_f^2 
e^{-\sqrt{2}\hbar k_f^2/(m \omega_\perp)}.
  \label{eq.Gamma0}
\end{equation}
The departure decreases  exponentially with the bias field $B_0$, as for a
usual time independent Ioffe trap~\cite{PhysRevA.56.2451}. However, 
in the modulated
trap, an additional exponential factor in $\omega/\omega_\perp$ reflects the fact
that the Floquet level is increased by one while the spin is flipped. This transition is
associated with the ``emission'' of a quantum of energy $\hbar\omega$, given to the
oscillating magnetic field.
Equation~(\ref{eq.Gamma0}) also shows that
the departure rate goes to zero for a modulation
frequency very close to the frequency 
$\mu B_0/\hbar+\omega_\perp/\sqrt{2}$, {\it i.e.} for 
vanishing $k_f$.
This cancellation is due to the fact that the coupling
term of Eq.~(\ref{eq.dRdt}) is odd in $x$ whereas the initial
state is even  and the final state, whose  wavevector
is vanishing, is flat.
 The departure rate towards the Floquet state $n_F=-1$ is identical
to Eq.~(\ref{eq.Gamma0}), $\omega$ being replaced by $-\omega$. 
Since we assumed
$\omega\gg \omega_\perp$, the loss rate towards the Floquet state $n_F=-1$ is
negligible compared to Eq.~(\ref{eq.Gamma0}).

\begin{figure}
\includegraphics{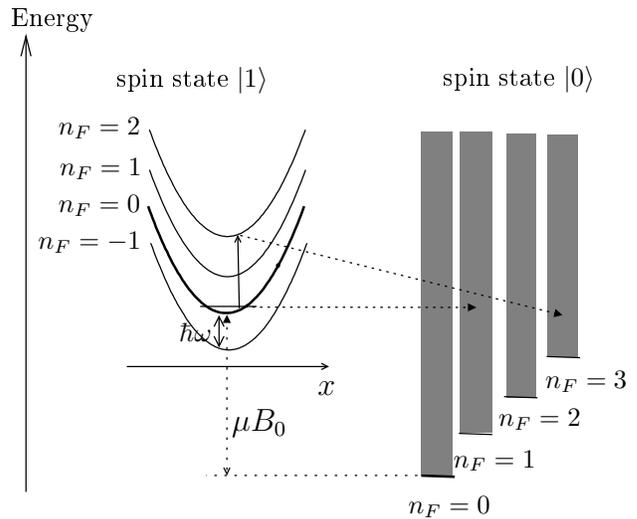}
\caption{Transitions responsible for spin-flip losses.
For the spin  state $\ket{1}$, the potential energy term of 
Eq.~(\ref{eq.Hadtrans}) is represented as well as the energy 
of the ground state in the $n_F=0$ manifold.
 For the spin state $\ket{0}$, 
we represented, for each Floquet manifold $n_F$, the whole energy spectrum, 
which is 
a semi-continuum starting at an energy $n_F \hbar\omega$.
The transitions induced by the term of Eq.~(\ref{eq.dRdt}) are
shown as dotted arrows whereas transitions due to the
term $H_2$ of Eq.~(\ref{eq.H2spinflip}) are shown as solid lines.
The initial state is the spin state $\ket{1}$ of Floquet number $n_F=0$.
For the two final Floquet states $n_F=1$ and $n_F=3$, only the
dominant processes are sketched, whose amplitudes are
$U_0$ and $U_1$ respectively, where $U_n$ is given in Eq.~(\ref{eq.Un}).
In this picture, the odd Floquet state of $\ket{0}$ the closest to
resonance corresponds to $n_F=3$ and losses are dominated by
$\Gamma_1$, where $\Gamma_n$ is given in Eq.~(\ref{eq.gamman}).
}
\label{fig.spinflipFloquet}
\end{figure}

The condition $\omega\gg \omega_\perp$ and  
Eq.~(\ref{eq.Gamma0}) show that the loss
rate is exponentially small when $\omega$ reaches $\omega_1=(\mu
B_0/\hbar+\omega_\perp/\sqrt{2})/3$, the value for which the initial state has the
same energy as the untrapped state of Floquet number $n_F=3$ and of vanishing
momentum. For modulation frequencies smaller than $\omega_1$, second order processes
resonantly couple the initial state to the untrapped state $\ket{0}$ of Floquet
number $n_F=3$. In theses processes, represented in Fig.\ref{fig.spinflipFloquet},
the term $H_2$ of Eq.~(\ref{eq.H2spinflip}) first transfers the atoms into 
the virtually
populated intermediate trapped state $\ket{1}$ of Floquet number $n_F=2$ before the
term $i\hbar \frac{d{\cal R}^{-1}}{dt}{\cal R}$ realizes the transfer to the
untrapped state $\ket{0}$ of Floquet number $n_F=3$. Although $H_2$ is weak, the
loss rate associated with the second order processes is much larger than the
exponentially small $\Gamma_0$.

 More generally, for a given modulation frequency,
losses are dominated by transitions towards untrapped states of Floquet number
$n_F=2n+1$ where $n=E((\mu B_0/\hbar +\omega_\perp/\sqrt{2})/(2\omega)-1/2 )$,
the function $E(x)$ being the integer part of $x$.
 Those transitions correspond to processes
where the atom is first brought to the intermediate state $\ket{1}$ of Floquet
number $n_F=2n$ by $n$ transitions produced by the term  $H_2$ and is then
transferred to the untrapped state $\ket{0}$ of Floquet number
$n_F=2n+1$ by 
the term $i\hbar d{\cal R}^{-1}/dt {\cal R}$ of Eq.~(\ref{eq.dRdt}).
 Perturbation theory  gives
an effective coupling between  the state $\ket{1}$ of Floquet number $n_F=0$ and the
states $\ket{0}$ of Floquet number $n_F=2n+1$ which is
\begin{equation}
U_{n}=-i\frac{\hbar \omega b'(x-iy)}{n!2\sqrt{2}B_0}
(m\omega_\perp^2(x^2+y^2)/(16\hbar\omega))^{n}.
\label{eq.Un}
\end{equation}
 The eigenstates of $H_0$
in the virtual intermediate states do not appear because,
since we assumed $\omega\gg \omega_\perp$,
the energy difference between the
intermediate states is negligible and a resummation is possible.

 The departure rate from the initial state
towards the Floquet state $n_F=2n+1$ is computed from $U_n$
using the Fermi Golden Rule. As for the calculation of $\Gamma_0$,
it is sufficient to compute the departure rate in the $x$ direction
and we obtain
\begin{equation}
\Gamma_{n}=\begin{array}[t]{l}
\frac{m\omega^2 b'^2}{8 \hbar B_0^2}
\frac{(m\omega_\perp^2)^{2n}}{n!^2(16\hbar \omega)^{2n}} \\
\left |
\int \!\!\int\! dxdy (x-iy)e^{ik_f x}\phi_0(x,y)(x^2+y^2)^n
\right |^2
\end{array}
\label{eq.gamman}
\end{equation}
where 
$k_f=\sqrt{2m(\mu B_0 + \hbar\omega_\perp/\sqrt{2}-(2n+1) \hbar\omega)}/\hbar$ is
the wavevector of the final state. 
Using the gaussian expression for the ground state  $\phi_0(x,y)$
in Eq.~(\ref{eq.gamman}), 
we can show that $\Gamma_n$ is
the product of a polynomial in $k_f$ and of the exponential factor
$e^{-\sqrt{2}\hbar k_f^2/(m\omega_\perp)}$.
 The minima of the polynomial correspond to destructive
interferences between the probability  amplitudes of paths having different
intermediate vibrational states.
 Since $\hbar k_f^2/m$ is reduced by $4\omega$
when increasing $n$ by one and since we assumed
$\omega\gg\omega_\perp$, the exponential factor ensures,
as stated above, that
the total loss rate is  dominated by the departure towards
the highest Floquet subspace.

\begin{figure}
\includegraphics{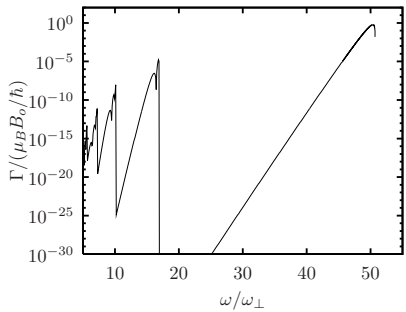}
\caption{Loss rate from the vibrational ground state versus the
modulation frequency. The
 longitudinal magnetic field and the quadrupole gradient
are chosen so that $\mu B_0/(\hbar\omega_\perp)=50$.}
\label{fig.losses}
\end{figure}

 Figure \ref{fig.losses} gives the departure rate of the
trapped ground state as a function of $\omega$ for
$\mu B_0=50 \hbar\omega_\perp$.
 We observe several peaks that reflect the resonance behavior at
integer values of $(\mu B_0+\omega_\perp/\sqrt{2}-\omega)/(2\omega)$.
 The height of the resonances goes down with
the integer $n$ as expected since the order of the
transition increases with  $n$. We verify that
the loss rate is dominated by the losses towards the
Floquet state of highest odd Floquet number, as expected.
Between two resonances, we observe the expected  exponential
decrease of the loss rate.
We observe a structure in the loss rate for
losses to Floquet state larger than one, as
expected.

 The lifetime of a thermal Maxwell-Boltzmann distribution
is obtained after averaging the loss rate
over the thermal distribution. In this calculation,
$\phi_0$ in  Eq.~(\ref{eq.gamman}) is replaced by
the eigenstate $\phi_{i}(x)\phi_{j}(y)$, where $i$ and
$j$ denotes the vibrational level.
 Neglecting changes of the final energy $\hbar^2k_f^2/(2m)$
across the thermal distribution, we can compute the departure
rate $\Gamma_0$ for a Maxwell-Boltzmann distribution.
 More precisely, writing Eq.~(\ref{eq.gamman}) as a fourth integral and using
properties of Wigner functions
we find, for temperatures $k_B T \gg \hbar \omega_\perp$, 
\begin{equation}
\Gamma_0=\frac{\pi^3}{2}\omega_\perp 
\frac{\hbar^3\omega_\perp\omega^2}{(k_B T)^3}
e^{-(\mu B_0 -\hbar \omega)/(k_B T)}.
\end{equation}
Because of the crude approximation that the final energy does not 
depend on the initial state, this result is only valid up to a factor 
of the order of unity. 

 Experimentally, spin-flip losses can easily be avoided by
properly choosing the modulation frequency. For example,
let us assume the transverse oscillation frequency of the
instantaneous trap at maximum current is
$\omega_\perp/(2\pi)=50~$kHz and the longitudinal
magnetic field fulfills $B_0=50\hbar\omega_\perp/\mu$.
 If $\mu=\mu_B$ where $\mu_B$ is the Bohr magneton,
this corresponds to $B_0=1.8~$G and the Larmor
frequency $\mu B_0/\hbar$ is 2.5~MHz.
 In these conditions, the loss rate is dominated by the term $\Gamma_0$ of
Eq.~(\ref{eq.Gamma0}) as long as the oscillation frequency
is larger than  0.84~MHz and, in this frequency range, it is
smaller than 1~s$^{-1}$ as soon as $\omega<2.2~$MHz. For $\omega<0.84~$MHz,
losses become  dominated by transitions towards states of
higher Floquet numbers and the loss rate is peaked at
modulation frequencies close to integer values of
$\mu B_0/(2\hbar \omega) +\omega_\perp/(2\sqrt{2}\omega)-1/2$.
 In particular, the loss rate goes up to  about 25~s$^{-1}$
for a modulation frequency  close to  0.8~MHz. Thus, the
vicinity of this resonance should be avoided experimentally.
 Resonances of higher order are less problematic since the
maximum loss rate they induce is smaller than  0.1~s$^{-1}$.

\section{Radio-frequency evaporation in the modulated guide}
 In this section we present general considerations on
forced evaporation in a modulated guide. 
Since evaporative cooling is most efficiently realized in a 3D trap,
longitudinal confinement is required.
A 3D trap can be obtained 
from the modulated guide of section II by applying a z-dependent 
constant longitudinal field $B_0(z)$.
 Here, we consider evaporation in the transverse plane ($xy$)
at a given $z$ position and denote as $B_0$ the longitudinal magnetic 
field.
For this
purpose, in addition to the previous trapping potential, we apply a weak radio
frequency magnetic field polarized in the x-direction, of 
frequency $\omega_{RF}$ and of amplitude $B_{RF}$. We
consider here an atom of magnetic moment $\mu {\bf J}/\hbar$, 
where ${\bf J}$ is the atomic spin angular momentum.

Let us first give simple predictions, that only rely  on the 
fact that, 
because of the modulation at $2\omega$ of the trapping potential, 
the atomic Larmor frequency is
modulated in time.  The modulation amplitude $\delta \Omega$
increases, in the transverse plane,
with the distance $r$ from the trap center according to
$\delta\Omega=\mu b'^2 r^2/(4\hbar B_0)$.
Considering only the internal atomic dynamics at a given 
position, the modulation of the Larmor frequency is equivalent, 
within the rotating wave approximation, to a frequency 
modulation of the radio-frequency field.
In this picture, the radio frequency  spectrum 
consists of a carrier at the frequency
$\omega_{\rm RF}$ and side-bands spaced by $2\omega$,
the relative
amplitude of the $n^{\rm{th}}$ sideband with respect to the carrier 
being $J_n(\mu (b'r)^2/(8B_0\hbar\omega))$, 
where $J_n$ is the Bessel function of the first
kind. 
Thus, for a given frequency of the applied RF field, the 
coupling to the untrapped
state is resonant for the positions $r_n$ such that
\begin{equation}
\omega_{RF}=\mu
B_0/\hbar+\mu b'^2 r_n^2/(4 B_0)-2n\omega,
\label{eq.rn}
\end{equation} where $n$ is a integer.
The coupling 
between the spin states 
close to a resonance position $r_n$ is 
\begin{equation}
V_n=V_0 J_n(\mu(b'r)^2/(8B_0\hbar\omega))
\label{eq.Vnfm}
\end{equation}
where $V_0$ is the coupling produced by the radio-frequency 
field in the absence of modulation.
Such a shell structure of  the  spin-flip
transition resonances is characteristic of 
AC magnetic traps. For example, the same behavior is expected 
in TOP traps~\cite{TOPres}, where the Larmor frequency 
is also modulated in time.

In  the following, 
we verify the statements made above by a more rigorous derivation.
As in the previous section,  we consider
the representation in which the spin up state points along 
the local instantaneous magnetic field.
The RF field produces a term in the Hamiltonian experienced 
by the atoms which is, to first order in the angles 
$\theta_x=-b'x\cos(\omega t)/B_0$ and 
$\theta_y=b'y\cos(\omega t)/B_0$,
\begin{equation}
\begin{array}{ll}
H_{RF}& = \mu B_{\rm RF}\cos(\omega_{\rm RF}t)
J_x
\\
  &-\mu B_{\rm RF} b'y/B_0
\cos(\omega_{\rm m}t)\cos(\omega_{\rm RF}t)J_z.
\end{array}
\end{equation}
The right
hand side is divided in two terms. 
The first term (first line) corresponds to the usual coupling
between the spin states  in the presence of the RF
field. The second term (second line) appears due to the time dependence of
$\mathcal{R}$.
 As in the previous section, in the following we consider the case
of a spin one state and we restrict ourself to the two spin states
$\ket{1}$ and $\ket{0}$. We then have 
$H_{{RF}_1}=H_{{RF}_1}+H_{{RF}_2}$,
where
 \begin{equation}
H_{{RF}_1} = \mu B_{\rm RF}\cos(\omega_{\rm RF}t)
(\ket{1}\bra{0}+\ket{0}\bra{1})/\sqrt{2}
\label{Hamiltonien_RF_1}
\end{equation}
and 
 \begin{equation}
H_{{RF}_2} = -\frac{\mu B_{\rm RF} b'y}{B_0}
\cos(\omega_{\rm m}t)\cos(\omega_{\rm RF}t)
\ket{1}\bra{1}.
\label{Hamiltonien_RF_2}
\end{equation}

To analyze the effects of the RF field, we use the Floquet
representation where two quantum number are used: the Floquet number
$n_F$ associated to the modulation frequency $\omega$ and
$N_{\rm RF}$, the number of radio-frequency  photons. We consider 
an atom in the trapped spin $\ket{1}$ state,
with $N_{\rm RF}$ radio frequency (RF) photons. Because
$B_{RF}$ is weak, we only consider transitions involving a single RF
photon and we only consider transitions to 
the quasi-resonant states where the spin is 0 and the number 
of RF photons is $(N_{\rm RF}-1)$.

\begin{figure}
\scalebox{0.45}{
\includegraphics{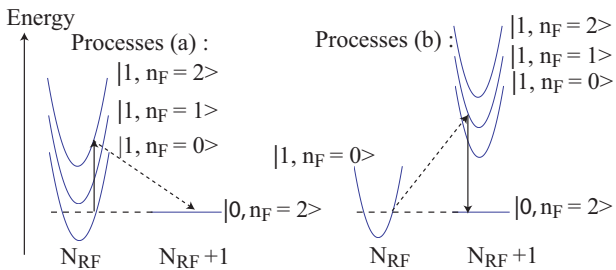}}
\caption{Processes involved in the transition from the 
trapped state $\ket{1,n_F=0,N_{\rm RF}}$ to the untrapped state 
$\ket{0,n_F=2,N_{\rm RF}-1}$. 
The processes (a) involves the Hamiltonian $H_2$ of
Eq.~(\ref{eq.H2spinflip}) (solid line)
and 
the term $H_{{RF}_1}$ of Eq.~(\ref{Hamiltonien_RF_1})
(dashed line). 
The process (b) involves the term 
$H_{{RF}_2}$ of Eq.~(\ref{Hamiltonien_RF_2}) (dashed line) 
and the 
term 
$i\hbar \partial {\cal R}^{-1}/\partial t {\cal R}$ 
of Eq.~(\ref{eq.dRdt}) (solid line).
 For $\omega\ll \mu B_0$, the processes (a) are the dominant one.}
\label{fig.RFprocesses}
\end{figure}

Let us suppose the initial trapped state is 
in the $n_F=0$ manifold. 
The  term 
$H_{{RF}_1}$ couples the initial state to the spin 0 state in the  
manifold $|N_{\rm RF}-1,n_F =0>$. 
This transition
is resonant for the position $r_0$ given by Eq.~(\ref{eq.rn})
and the coupling to the spin 0 state is $\mu B_{RF}/\sqrt{2}$. 
The initial state can also be transferred to 
the spin 0 state in the $|N_{\rm RF}-1,n_F =\pm 2>$
manifolds by higher order processes. These transitions,  resonant
 for 
the position $r_{\pm 1}$ given by Eq.~(\ref{eq.rn}), 
can occur via two kinds of processes, 
represented in Fig.\ref{fig.RFprocesses} in the case where the final state 
lies in the manifold $n_F=2$.
In the first  process, $H_2$ of Eq.~(\ref{eq.H2spinflip}) 
couples the initial
state to the spin 1 state in the manifold $|N_{\rm RF},n_F=\pm 2\rangle$, 
which is then transferred
by the  term $H_{{RF}_1}$ to the spin 0 final state (process a). In the
second  kind of processes (process b), 
the transfer from the spin state 1 to the
spin state 0 is ensured by the
term $i\hbar
\frac{d\mathcal{R}^{-1}}{dt}\mathcal{R}$ 
of the Hamiltonian (see Eq.~(\ref{eq.dRdt})), and the 
term $H_{{RF}_2}$ of the radio-frequency coupling is involved.
In the
case where $\omega\ll \mu B_0/\hbar$, the process (b) have a negligible
amplitude and only the process (a) is important.

In a more general way, the initial state can be transferred to the
final state of odd Floquet number $n_F=2n$, the transitions 
being resonant at the positions $r_n$ given by Eq.~(\ref{eq.rn}). 
The dominant processes
involve the first term $H_{RF}$ given in Eq.~(\ref{eq.rn}) and the term
$H_2$ to order $n$. The effective coupling
between the trapped state and the spin 0 state of Floquet number
$2n$, computed to lowest order in $H_2$, is
\begin{equation}
V_{n,eff}=\frac{(-1)^n}{n!} \left
(\frac{\mu b'^2r^2}{16B_0\hbar\omega}\right )^{|n|} 
\frac{\mu B_{RF}}{\sqrt{2}}.
\label{eq.Vneff}
\end{equation}
 We recover here the result of Eq.~(\ref{eq.Vnfm}), in the 
limit considered here  where $\hbar\omega\gg{\mu b'^2r^2}/(8B_0)$. Thus 
the simple description in terms of  
frequency modulation of the Larmor frequency is sufficient to describe
the physics.

In conclusion, we have shown that the radio-frequency field is
resonant
for  different trap locations, whose 
potential energy differ by $2\hbar \omega/k_B$.  For a modulation
frequency of $50$~kHz, the potential energy difference between two
resonances is $3~\mu$K. For a temperature of the order of $3~\mu$K or
higher, some resonances are present inside the atomic cloud and induce
spin-flip losses.
 To overcome this problem, a precooling stage in a static trap down to
temperatures smaller than $3~\mu$K is required. 
For clouds whose temperature is smaller than $3~\mu$K, 
the evaporation process in the modulated guide 
involves only one radio-frequency knife, so that
the evaporative cooling is similar to that realized in  a 
trap made by DC currents. 
Choosing the frequency of the radio frequency
field so 
that the transition involved in the cooling process is the transition 
that does not change the Floquet number
is interesting for two reasons. 
First, as shown in Eq.~(\ref{eq.Vneff}), the coupling between the trapped 
and the untrapped state of this transition 
is larger than that of higher order transitions 
that change the Floquet number.  
 Second, this coupling is  
homogeneous and 
is thus constant when $\omega_{RF}$ is chirped.

\section{Conclusion}
 The careful study  of the limitations of the modulation
technique to smooth  wire guide roughness performed in this article
show that this technique is very robust and accepts a wide range of
modulation frequencies. More precisely, on one side, we have shown that the
unwanted effects of the modulation on the longitudinal motion are
negligible for realistic parameters as soon as modulation frequency
is larger than 10~kHz: both the heating of the cloud and the
remaining effective roughness are very small.
 On the other side, the calculation of the losses due to
spin-flip transitions shows that, for realistic parameters,
these losses are negligible as soon as the modulation frequency is
smaller than a few hundred of kHz.
 The modulation technique is thus a very promising tools that should enable
to fully take advantage of atom chips devices. In particular, the
study of one-dimensional gases
in the strong interacting regime~\cite{tolra:190401,Parades,Weiss1D} 
on an atom chip can be considered.

 The smoothing technique studied in this paper 
may be used in any situations where a rough potential is proportional 
to a quantity that can be modulated, so that  
the calculations developped in section \ref{sec.long} may
apply to other physical systems.

\section{Aknowledgment}
The authors thank K. M{\o}lmer for helpful discussions.
The Atom Optic group of Laboratoire Charles Fabry is part of 
the IFRAF institute. This work was supported by the EU under the grants 
No MRTN-CT-2003-505032 and IP-CT-015714.

\appendix
\section{Derivation of the energy exchange rate}
The derivation of the heating rate follows that of the Fermi Golden rule.
For the calculation, we assume
a quantification box of size $L$ and periodic boundary
conditions.
We assume the atom is initially in the state of momentum
$p_0$ in the Floquet subspace $n_F=0$. For simplicity we
consider only the transitions towards the Floquet state $n_F=-1$.
 After a time $t$ much smaller than the departure rate,
the change in kinetic energy $\Delta E$ can be deduced from perturbation
theory and we obtain
\begin{equation}
\Delta E= \sum_q |u_q|^2 f(q,t).
\label{eq.DeltaE}
\end{equation}
Here $u_q=\int dz u(z)e^{iqz}/L$ is the Fourier component of
wavevector $q$ of $u(z)$ and
\begin{equation}
f(q,t)=(\epsilon-\hbar\omega)
\frac{\sin^2(\epsilon t/2)}
{\epsilon^2},
\end{equation}
where $\epsilon=\hbar\omega+\hbar^2q^2/(2m)+\hbar p_0 q/m$
is the energy change associated to the transition involving
the Fourier component of $u$ of wavevector $q$.
The terms $u_q$ are complex random numbers without correlation between
them and of mean square value $\langle |u_q|^2\rangle=S(q)L/(2\pi)$
where $S(q)=1/(2\pi)\int dz e^{iqz} \langle u(z)u(0)\rangle$
is the spectral density of $u$.
 For a large enough quantification box, the term
$f(q)$ barely changes
between adjacent  Fourier components and one can replace
$\sum_q |u_q|^2 f(q)$ by $\int dq S(q) f(q)$ in Eq.~(\ref{eq.DeltaE}).
For a time $t$ large enough so that the function
$(\epsilon-\hbar\omega)S(q(\epsilon))$
is about constant on the interval $\epsilon\in[-\hbar/t,\hbar/t]$,
the  term $\sin^2(\epsilon t/2)/\epsilon^2$
can be replaced by the distribution
$t\delta(\epsilon)\pi/2$. We then recover the first term of
Eq.~(\ref{eq.dEdtquant}).
The previous condition on $t$ and the condition that $t$ is
much smaller than the departure rate $\Gamma$ can be fulfilled simultaneously
only if the function
$(\epsilon-\hbar\omega)S(q(\epsilon))$ is about constant on
the interval $\epsilon\in[-\hbar\Gamma,\hbar\Gamma]$. This is the
condition of the Markovian approximation. This condition
is fulfilled provided that both $S(q)$  and its correlation
length are  small enough.

 The calculation is similar for losses towards the Floquet manifold
$n_F=1$ and one finally recover Eq.~(\ref{eq.dEdtquant}).
The Markovian condition for the transition towards
the Floquet manifold
$n_F=1$ is not fulfilled for
initial momentum very close to $\sqrt{2m\hbar \omega}$ since the
atoms are then sensitive to the fact that the continuum is not
infinite for $\epsilon <0$. Similar non Markovian situations 
have been studied, for example, in photonic band gap
materials~\cite{NonMarkovian}
and oscillatory behavior and decay towards a non vanishing 
population of the initial state are expected.

\section{Adiabatic potential in the dressed state representation}
 In this appendix, we rederive the adiabatic potential given Eq.~(\ref{eq.Had})
using a dressed representation, where a local z-dependent
unitary transformation ${\cal O}(z)$
is applied to the Floquet states so that the resulting dressed states
 $\ket{n}(z)$
are eigenstates of the potential energy part of
the Floquet Hamiltonian (term 
$\hbar \omega n_F$ of $H_0$ given in Eq.~(\ref{eq.H0chauff}) and 
term $H_2$ of Eq.~(\ref{eq.H2chauff})).
By symmetry, the energy of the dressed states  $\ket{n}(z)$
is  $n\hbar\omega$, as that of the bare Floquet states.
Using the properties of the Bessel functions
$(J_{k+1}(x)+J_{k-1}(x))x/(2k)=J_k(x)$
and $\sum_n J_n(x)^2=1$, we show that
 the decomposition of $\ket{n}(z)$ in the undressed Floquet
basis $(\ket{k}_0)$ is
\begin{equation}
\ket{n}(z)=\sum_{k=-\infty}^{\infty} J_{k}(u(z)/\omega) \ket{n+k}_0={\cal
  O}(z)\ket{n}_0.
\label{eq.ndressed}
\end{equation}
 This well-known result has been used in several other
situations~\cite{devbessel}.
 In the dressed state representation, the
state of the system is $\tilde{\psi}={\cal O}^{-1}\psi_0$
where ${\psi}_0$ is the state of the system in
the undressed representation and
the momentum operator,
$\tilde{p}={\cal O}^{-1}p{\cal O}$, is
\begin{equation}
\tilde{p}= p - \sum_{n,k}
{\bra{k,z}}i\hbar\partial_z\ket{n,z}
\ket{k}\bra{n}
\end{equation}
where $p=-i\hbar\partial_z$
is the momentum operator that preserves the Floquet number and
$\partial_z$ is a short notation for $\partial/\partial_z$.
 Thus, in the dressed state representation, the Hamiltonian is decomposed
into three terms:
\begin{equation}
\tilde{H}_0=p^2/2m -n_F \hbar\omega
\end{equation}
that does not couple different Floquet states,
\begin{equation}
\tilde{H}_1=-\hbar/(2m)
\begin{array}[t]{l}
\left (
p\sum_{n_1,n_2}\ket{n_1}\bra{n_1}i\partial_z\ket{n_2}\bra{n_2}
\right .\\
\left . +\sum_{n_1,n_2}\ket{n_1}\bra{n_1}i\partial_z\ket{n_2}\bra{n_2}p
\right ),\\
\end{array}
\end{equation}
and
\begin{equation}
\tilde{H}_2
\begin{array}[t]{l}
=-\hbar^2/(2m)\displaystyle\sum_{n_1,n_2,n_3}\ket{n_1}\bra{n_2}
\bra{n_1}\partial_z\ket{n_3}
\bra{n_3}\partial_z\ket{n_2}\\
\end{array}.
\end{equation}
 Since $J_k'=(J_{k-1}-J_{k+1})/2$ and
$\sum_k J_k J_{k+n}=\delta_n$, $\tilde{H}_1$ couples
adjacent Floquet states. This term is responsible for the
heating of the cloud.
 On the other hand, $\tilde{H}_2$ contains a term $\tilde{H}_{2,ad}$ that does
not change the Floquet number. Using the above properties
of the Bessel function, we find that $\tilde{H}_{2,ad}$ is
just the adiabatic potential of Eq.~(\ref{eq.Had}).


\begin{thebibliography}{17}
\bibitem{noteTOP}{These two requirements are also needed 
for the TOP trap to work.}
\end{thebibliography}
\end{document}